\documentclass[10pt]{article}
\usepackage{amssymb}
\usepackage{graphicx}
\usepackage{layout}
\usepackage{bm}

\begin{document}

\author{Matteo Luca Ruggiero$^\S$
\\ \\
\small
$^\S$ Dipartimento di Fisica, Politecnico di Torino and INFN, Sezione di Torino\\
\small E-mail: matteo.ruggiero@polito.it }
\title{Rotation Effects and The Gravito-Magnetic Approach}

\maketitle

\begin{abstract}
Gravito-electromagnetism is somewhat ubiquitous in relativity. In
fact, there are many situations where the effects of gravitation
can be described by formally introducing  "gravito-electric" and
"gravito-magnetic" fields, starting from the corresponding
potentials, in analogy with the electromagnetic
theory\cite{mashhoon03},\cite{ruggiero02} (see also A. Tartaglia's
contribution to these proceedings). The "many faces of
gravito-electromagnetism"\cite{jantzen92} are related to rotation
effects  in both approximated and full theory approaches. Here we
show that, by using a 1+3 splitting, relativistic dynamics can be
described in terms of gravito-electromagnetic (GEM) fields in full
theory. On the basis of this formalism, we introduce a
"gravito-magnetic Aharonov-Bohm effect", which allows to interpret
some rotation effects as gravito-magnetic effects. Finally, we
suggest a way for measuring the angular momentum of celestial
bodies by studying the gravito-magnetic effects on the propagation
of electromagnetic signals.
\end{abstract}

\maketitle


\section{Gravito-Magnetic Aharonov-Bohm Effect}
Let a physical reference frame (PRF) be defined by a time-like
congruence $\Gamma$ of world-lines of particles constituting the
3-dimensional reference frame.\cite{rizzi_ruggiero04} The motion
of free particles relative to $\Gamma$ can be described by
projecting the  equation of motion onto the 3-space of the PRF by
means of the "natural splitting".\footnote{We refer to the
projection technique developed by Cattaneo: see
\cite{rizzi_ruggiero04} and references therein.} The projected
equation  can be written in the form\cite{rizzi_ruggiero04}
\begin{equation}
\frac{\hat{D}\widetilde{p}_{i}}{dT}=m \widetilde{E}^{G}_i+m\gamma
_{0}\left( \frac{\bm{\tilde{v}}}{c}\times
\bm{\tilde{B}}_{G}\right) _{i} \label{eq:lorentzgem}
\end{equation}
in terms of the GEM fields $\tilde{\bm{E}}_G, \tilde{\bm{B}}_G$.
In other words, the dynamics of particles, relative to the
reference frame $\Gamma$, is described in terms of motion under
the action of a GEM Lorentz force, which is obtained without any
approximations on fields and velocities. On the basis of the the
analogy between the "classical" Lorentz force and the GEM Lorentz
force (\ref{eq:lorentzgem}), we may formally introduce the
gravito-magnetic Aharonov-Bohm phase shift and the corresponding
time delay\cite{rizzi_ruggiero04}
\begin{equation}
\Delta T = \frac{2\gamma _{0}}{c^3}\oint_{C}\widetilde{\bm{A}}_{G}\cdot {\rm d}%
\widetilde{\bm{x}}=\frac{2\gamma
_{0}}{c^3}\int_{S}\bm{\widetilde{B}}_{G}\cdot {\rm
d}\bm{\widetilde{S}} \label{eq:deltatau1}
\end{equation}
Eq. (\ref{eq:deltatau1}) describes the time delay for two massive
or massless beams, which propagate in opposite directions along a
closed path, with the same velocity in absolute value. Because of
the gravito-magnetic field, the two beams take different times (as
measured by a clock at rest in $\Gamma$) for a round trip, and
(\ref{eq:deltatau1}) expresses this time delay. In particular,
some rotation effects can be thought of as gravito-magnetic
Aharonov-Bohm effects: this is the case, for instance, of the
Sagnac effect, both in flat and curved-space time.
\cite{rizzi03a}, \cite{ruggiero_elba04}.

\begin{figure}[top]
\includegraphics[width=12cm,height=5cm]{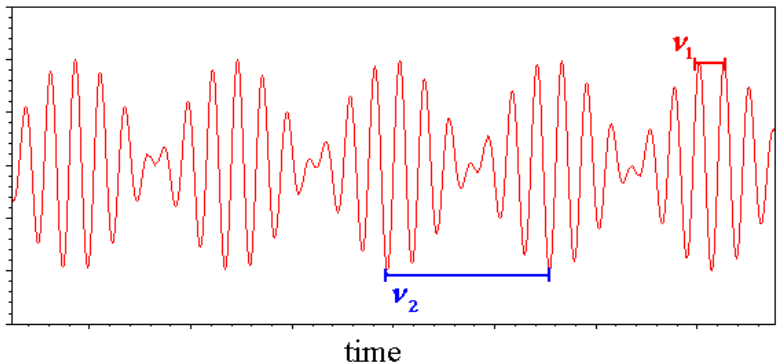}
\caption{ \small The basic signal frequency $\nu_1$ depends on the
mass of
 the rotating body only, while the amplitude modulation frequency $\nu_2$ depends on its  angular
momentum only.} \label{fig:beats1}
\end{figure}

\section{Measuring the Angular Momentum of Celestial Bodies}

Gravito-magnetic effects originating from masses rotation can be
exploited for measuring the angular momentum of celestial bodies.
Shapiro \textit{et al.}
\cite{shapiro71},\cite{marte1},\cite{marte2} verified the
influence of the solar mass on signals propagation, however, in
weak field approximation, the time of flight does depend also on
the angular momentum, and its contribution has the
form\cite{tartaglia04}
\begin{equation}
t_{J} = \pm \frac{2GJ}{c^4b}f  \label{eq:angmom1}
\end{equation}
where $J$ is the magnitude of the angular momentum, whose
direction is supposed to be perpendicular to the plane of
propagation of the signals; $b$ is the impact parameter and $f$ is
a geometrical factor. The double sign means that the propagation
time is shortened on one side of the rotating mass and lengthened
on the other side. The interplay of the motions of the source of
electromagnetic signals, the rotating mass and the observer
results in a change of the impact parameter in (\ref{eq:angmom1}):
this induces a frequency shift\cite{tartaglia04}. To fix the
ideas, let us consider electromagnetic signals coming from a
distant source (such as a star, or a pulsar); these signals, after
propagating in the field of the Sun, are received by an
Earth-based observer. For the sake of simplicity, let us suppose
that the source, the Sun and the observer are in the same plane.
Since the source is much more distant from the Sun than the
receiver, to first approximation  the time variation of the impact
parameter is due to the motion of the receiver only. Around the
occultation of the source by the Sun, the impact parameter depends
linearly on time and its time variation is approximately given by
$|\dot{b}|=v_0$, where $v_0$ is the apparent velocity of the
source in the sky. If the signal is emitted with a frequency
$\nu$, it is received with a (slowly varying) frequency
$\nu+\delta \nu$, such that\cite{tartaglia04}
\begin{equation}
\frac{\delta \nu }{\nu }=\allowbreak 4\frac{GM
}{c^2b}\frac{v_{0}}{c}-\frac{GM b}{c^2R^{2}} \allowbreak
\frac{v_{0}}{c}\allowbreak
 \mp 4\frac{GJ}{c^3 b^{2}}%
\frac{v_{0}}{c}  \label{eq:rel_freq_shift}
\end{equation}
where $R$ is the distance of the observer from the rotating body
whose mass is $M$. The order of magnitude of this frequency shift,
for a Solar System experiment, is\cite{tartaglia04}
\begin{equation} \frac{GM
}{c^2b}\frac{v_{0}}{c} \sim 10^{-10} \ \ \ \ \frac{GM b}{c^2R^{2}}
\allowbreak \frac{v_{0}}{c} \sim 10^{-14}  \ \ \ \ \ \frac{GJ}{c^3
b^{2}} \frac{v_{0}}{c} \sim 10^{-16}
  \label{eq:num_estimates}
\end{equation}
The frequency shift due to the mass has been recently measured by
Bertotti \textit{et al.}\cite{bertotti03}, using radio pulses from
Cassini spacecraft. The angular momentum contribution could be
measured by producing interference between the received signals.
In other words, if we record the signal before and after the
occultation, and then superimpose them, we get a beating function
(see Figure \ref{fig:beats1}). The beating has a basic
frequency\cite{tartaglia04}
\begin{equation}
\nu_1=\nu \left( 4\frac{GM }{c^2b}\frac{v_{0}}{c}-\frac{GM
b}{c^2R^{2}} \allowbreak \frac{v_{0}}{c} \right) \label{eq:freq1}
\end{equation}
which depends on the mass only, and a frequency of amplitude
modulation
\begin{equation}
\nu_2= \nu \left(4\frac{GJ}{c^3 b^{2}}%
\frac{v_{0}}{c} \right) \label{eq:freq2}
\end{equation}
which depends on the angular momentum only: the two effects are
then decoupled.  Though small, this rotation effect is, in
principle, detectable, and it would provide a way for measuring
the angular momentum of celestial bodies. More favourable
conditions for observing this and other gravito-magnetic effects
could be attained by studying the recently observed double pulsar
system J0737-3039.\cite{lyne04}


\end{document}